\documentclass[amsmath,amssymb, aps, onecolumn,superscriptaddress]{revtex4-2}
\usepackage[utf8]{inputenc}
\usepackage{hyperref}
\hypersetup{
	colorlinks=true,
	linkcolor=blue,
	filecolor=blue,
	citecolor = blue,      
	urlcolor=cyan,
}

\usepackage{amsmath}
\usepackage{amsfonts}
\usepackage{MnSymbol}
\usepackage{xcolor}
\usepackage{graphicx}
\usepackage{siunitx}

\renewcommand{\v}[1]{\textbf{#1}}

\newcommand{\Z}[1]{\mathbb{Z}}
\newcommand{\etal}{\textit{et al.}}

\newcommand{\jp}[1]{{\color{black}{#1}}}   

\usepackage{orcidlink}

\begin{document}

\title{Neutral atom quantum computing for materials science and quantum chemistry}
\author{J. D. Pritchard\,\orcidlink{0000-0003-2172-7340}}
\email{jonathan.pritchard@strath.ac.uk}
\affiliation{Department of Physics and SUPA, University of Strathclyde, Glasgow G4 0NG, UK}

\begin{abstract}
Neutral atom arrays have emerged as versatile platforms for performing both digital and analogue quantum computing and simulation, with demonstrations ranging from large-scale programmable Hamiltonians realising topological phases or weighted graph optimisation to error-corrected logical qubits with transverse gate operations. This paper provides a broad overview to the neutral atom platform, and the potential applications relevant to materials science and quantum chemistry.
\end{abstract}
\maketitle

\section{Introduction}
Quantum computing is predicted to have significant impact on areas of material science and quantum chemistry by providing hardware that can efficiently simulate quantum systems \cite{camino26a}, overcoming the classically intractable exponential scaling associated with complex molecules \jp{and materials} \cite{reiher17,mcardle20,liu22,nguyen26,santagati24}. Broadly, these problems reduce to finding ground state configurations of a given system, or modelling the temporal response to a perturbation by performing simulations of its dynamic evolution.

Digital quantum computing approaches to simulating material and chemistry problems have relied on algorithms including the variational quantum eigensolver (VQE) for finding ground states from which equilibrium configurations or correlation functions can be evaluated \cite{peruzzo14,stanisic22,tilly22}, or time dynamics simulations (TDS) which utilise Trotterisation \cite{suzuki90} to break time evolution into circuits that discretely approximate a continuous time-step \cite{lloyd96}, enabling simulation of molecular energies \cite{omalley16} and fermionic models \cite{barends15}. 

Recent studies exploring resource requirements for real materials of interest, such as Li$_2$CuO$_2$ used in battery cathodes, show practical quantum computation requires hundreds to thousands of qubits and gate depths of up to $10^4$ \cite{clinton24}. These approaches currently lie beyond the realm of noisy, intermediate-scale quantum (NISQ) hardware \cite{preskill18} and will require large, error-corrected platforms to achieve utility-scale computation. 

Despite these challenges, there is active development of low-overhead algorithms for quantum simulation of materials \cite{babbush18} with direct application to models of high-temperature superconductors \cite{kan25}, simulation of condensed-phase correlated electrons \cite{kivlichan20}, \jp{thermodynamic sampling of disordered materials \cite{camino25},} and in quantum chemistry applications such as nitrogen fixation \cite{lee21} and catalysis \cite{vonburg21}.

As an alternative to digital quantum circuits, quantum hardware can also be used to explore analogue approaches to quantum computation. These include quantum annealing to find ground state configurations, or quantum simulation under a programmable Hamiltonian to study dynamics. For the annealing paradigm \cite{farhi00,farhi01}, it is possible to map a material science or chemistry problem onto an intermediate representation such as a quadratic unconstrained binary optimisation (QUBO) problem \cite{glover22}. Relevant examples include crystal structure prediction \cite{gusev23}, finding the energy of defective graphene structures \cite{camino23}, finding optimal pathways in chemical reaction networks \cite{mizuno24} and molecular docking \cite{pandey22,yanagisawa24,triuzzi25}. This approach enables users to find ground state solutions by annealing a quantum system to prepare qubits in the lowest energy configuration corresponding to the QUBO solution, with demonstrations of this approach using fast annealing of superconducting qubits \cite{teplukhin21}, which can be extended to enable studies of quantum critical spin-glass dynamics \cite{king23,king25}. More generally however, state-of-the-art quantum simulations of programmable Hamiltonians are already in a regime that goes beyond current classical simulation methods whilst also out-performing current NISQ digital quantum computers \cite{daley22}.

Whilst a wide range of hardware technologies are currently under development for creating programmable and fault-tolerant hardware, neutral atoms have emerged as a highly scalable and versatile platform for quantum computing and simulation  \cite{saffman10,adams20,morgado21,pritchard26}. By generating arrays of individually trapped atoms, it is possible to create registers of over a thousand qubits \cite{manetsch25,chiu25,wang25} which can be coupled to highly excited Rydberg states to create strong, long-range interactions to realise high-fidelity two-qubit gates for circuit based quantum computing \cite{jaksch00,lukin01,isenhower10}. Recent advances in gate fidelity \cite{evered23,ma23, tsai25,tao26, evered26}, combined with the ability to dynamically reconfigure qubits whilst maintaining coherence \cite{bluvstein24} have lead to pioneering demonstrations of quantum error correction and transverse operations with logical qubits \cite{bluvstein24,rodriguez25,bluvstein25, zhang25, reichardt25, bedalov24}. The same platform can also be used to realise programmable Hamiltonians for continuous-time quantum simulation \cite{browaeys20}, enabling exploration of Ising systems in 1D and 2D \cite{bernien17,scholl21,ebadi21} including preparation of topological spin liquids \cite{semeghini21}, or using an $XY$ Hamiltonian to demonstrate symmetry protected topological phases \cite{leseleuc19a}. By encoding classical optimisation problems onto the Ising Hamiltonian, these platforms further enable solution of hard classical problems \cite{kim23,dalyac24,wurtz24} including maximum (weighted) independent set (MWIS) \cite{pichler18,ebadi21,kim24,oliveira25}, QUBO \cite{byun24,oliveira26}, factorisation \cite{park24} and graph colouring \cite{angkhanawin25}. This analogue approach has been exploited to perform sampling equilibrium solvent water molecule configurations within proteins \cite{darcangelo24} as well as providing heuristic methods for solving molecular docking problems \cite{garrigues25}.

In this review we explore applications of tweezer-based neutral atom platforms to problems relevant for materials and molecular sciences, exploring application of both analogue and digital approaches to quantum computing. In Sec.~\ref{sec:na} we introduce the neutral atom platform and key concepts, before in Sec.~\ref{sec:analogue} exploring applications of analogue quantum computing on neutral atom arrays including quantum simulation and classical graph optimisation approaches. Section~\ref{sec:qec} explores advances in digital quantum computing including progress towards fault-tolerant logical operations essential, before finally providing an outlook towards development of native fermionic quantum processors in Sec.~\ref{sec:fermion}.



\section{Neutral Atom Quantum Computing}\label{sec:na}
\begin{figure}[t]
\centering
\includegraphics[width=12cm]{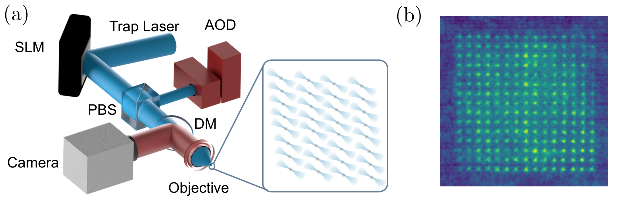}
\caption{(a) Schematic of a neutral atom quantum computer. Atom arrays are formed using arrays of microscopic optical tweezers created using reconfigurable holograms on a spatial light modulator (SLM) with atom sorting achieved using a fast mobile tweezer created using a crossed acousto-optical deflector (AOD) which can be combined on a polarising beam splitter (PBS). A high numerical-aperture objective lens is used to focus the tweezers and collect single atom fluorescence onto a camera for atom readout, using dichroic mirror to separate the wavelengths (DM). (b) An example image of atoms trapped in a $15\times15$ array using the experimental platform from \cite{nikolov23}. \label{fig:array}}
\end{figure}

Neutral atom quantum computers utilise individually trapped single atoms that are created by combining techniques of laser cooling with arrays of microscopic optical tweezers \cite{weiss17} shown schematically in Fig.~\ref{fig:array}. Tweezers arrays are created using acousto-optic deflectors \cite{endres16,lester15}, micro-lens arrays \cite{schaffner20}, or via holographic techniques using a spatial light modulators (SLMs) \cite{nogrette14,barredo16} that permit up to around $10^4$ reconfigurable tweezers \cite{manetsch25}. For larger-scale systems, meta-surface traps offer higher power handling to permit systems with $>10^5$ atoms in the future \cite{holman26}, with a major advantage over competing technologies being that the system retains a high degree of parallelism of operations as the qubit number is increased.

Atoms are trapped in tightly focused tweezers with waists around 1~$\mu$m to provide strong axial confinement. This enables stochastic single atom loading due to light-assisted collisions \cite{schlosser01}, which causes ejection of atoms pairs from the trap resulting in 50\% probability of loading either a single-atom or no atoms. In order to perform useful computation, it is necessary to subsequently assemble defect-free arrays, which would otherwise only occur with a probability of $1/2^N$ for an $N$ atom array. This is achieved by imaging the randomly loaded atomic array, and then using an additional fast, steerable atomic tweezer to rearrange the atoms in real-time \cite{barredo16}. 

To generate couplings between qubits, atoms are optically excited to high principal quantum number Rydberg states which experience strong, long range dipole-dipole interactions. For a pair of atoms separated by distance $r$, the dipole-dipole coupling typically results in a van der Waals interaction that scales as $V(r)=C_6/r^6\propto n^{11}$ \cite{saffman10}, meaning the relative strength can be controlled both by geometry (affecting $r$) and choice of principal quantum number ($n$). These strong interactions lead to an energy shift in the doubly-excited Rydberg state $\vert rr\rangle$, and for $V(r)>\Omega$, where $\Omega$ is the Rabi frequency of the optical coupling from the ground to Rydberg state, creates a \textit{dipole blockade} in which only a single Rydberg state can be created within a blockade radius $r_\mathrm{b}=\sqrt[6]{\vert C_6\vert/\Omega}$ \cite{lukin01}. As an example, for a pair of cesium atoms excited to the $80s_{1/2}$ Rydberg state with $\Omega/2\pi=1$~MHz, the blockade radius is $r_\mathrm{b}=12~\mu$m.

In the context of digital quantum computing, dipole blockade can be exploited on pairs of atoms to perform controlled gate operations, with the native gate being a controlled phase gate ($CZ$) \cite{jaksch00,lukin01}, which can be efficiently realised using global driving of atom pairs \cite{levine19} to perform parallel gate operations with further improvements in gate fidelity enabling $\mathcal{F}\gtrsim99.5\%$ \cite{evered23,ma23, tsai25,tao26, evered26} by implementing robust, ime-optimal pulses \cite{jandura22}.

Alternatively, for analogue operations the Rydberg interaction can be used to provide a programmable Ising Hamiltonian of the form

\begin{equation}
H_\mathrm{Ryd}/\hbar = \frac{\Omega}{2}\displaystyle\sum_i \sigma_i^x-\displaystyle\sum_i \Delta_i n_i + \displaystyle\sum_{j>i} V(r_{ij}) n_in_j ,\label{eq:ryd}
\end{equation}

where $\sigma_i^x=\vert r \rangle_i\langle g\vert+\vert g \rangle_i\langle r\vert$ is the coupling from the ground state $\vert g \rangle_i$ to the Rydberg state $\vert r \rangle_i$ on atom $i$, $\Delta_i$ is the detuning of the driving laser from the Rydberg state, $n_i=\vert r \rangle_i\langle r\vert$ and $r_{ij}=\vert r_i-r_j\vert$. In this regime, the system can be programmed by choice of geometry (enabled by the combination of arbitrary holographic arrays and defect-free loading), global excitation laser parameters and control of the on-site detuning using local light shifts. Below we will explore how this Hamiltonian can be used to perform quantum simulation or quantum optimisation on programmable atom arrays.

\section{Analogue Quantum Computing}\label{sec:analogue}

\subsection{Quantum Simulation using Atom Arrays}\label{sec:qsim}
Quantum simulation in neutral atom arrays has enabled rich exploration and discovery of new physics, even in systems as simple as a 1D chain of atoms with open boundary conditions. Early experiments demonstrated the ability to prepare ground states of up to 51 atoms in an Ising chain using only global control ($\Delta_i=\Delta$) of the Rydberg excitation laser \cite{bernien17}. Starting with atoms spaced by distance $a$ and initialised in $\vert g\rangle$, the Rydberg laser is ramped to $\Omega$ with a large negative detuning ($\Delta<0$) for which the ground state of the many-body system overlaps with the initial state $\vert g\rangle$. The detuning $\Delta(t)$ is then adiabatically ramped to a positive value ($0<\Delta<V(a)$) to prepare the array in the ground state of the interacting system where $V(a)=C_6/a^6$ defines the energy associated with the dipole-interaction between neighbouring sites.

\begin{figure}[t]
\centering
\includegraphics[width=12cm]{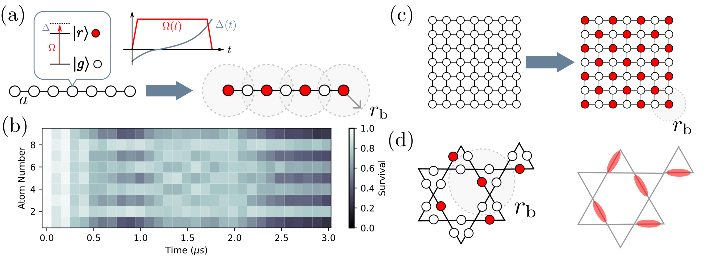}
\caption{Quantum simulation on a neutral atom array (a) A 1-D chain of atoms spaced with $a<r_\mathrm{b}$ can be prepared in the $\mathbb{Z}_2$-ordered phase with every other site Rydberg excited using adiabatic control of the global drive parameters $\Omega(t), \Delta(t)$ (b) Experimental data showing adiabatic evolution for a chain of 7 atoms towards the  $\mathbb{Z}_2$ ordered state adapted from \cite{oliveira25}. (c) Extending to 2D geometries provides access to phase-diagrams on square lattices \cite{scholl21,ebadi21} (d) Placing atoms on the edges of a kagome lattice enables mapping to a dimer model on which it has been possible to prepare topological quantum spin liquids on a neutral atom array \cite{semeghini21}. \label{fig:sim}}
\end{figure}

The resulting state is dependent upon the final value of $\Delta$ and ratio $r_\mathrm{b}/a$, where $a$ is the lattice spacing. For an odd chain length $N$ with $a=0.6r_\mathrm{b}$, the system prepares a $\Z_2$-ordered crystalline state with every other site Rydberg excited as illustrated in Fig.~\ref{fig:sim}(a) and (B), with higher order periods accessible by increasing the ratio of $r_\mathrm{b}/a$. This system provides a powerful testbed for investigating quantum critical dynamics, enabling experimental verification of the quantum Kibble-Zurek mechanism \cite{polkovnikov05,zurek05,dziarmaga05} by studying critical exponents extracted by crossing the phase transition to the $\Z_2$-ordered phase at different speeds and counting excitations via domain wall formation \cite{keesling19}. Within these experiments, when quenching the system to resonance ($\Delta=0$) following preparation of the $\Z_2$-ordered phase, the system demonstrates coherent and persistent oscillation of the crystalline order \cite{bernien17}, providing the first experimental evidence for many-body quantum scars offering a new class of quantum dynamics \cite{turner18}. This rich system provides a useful testbed to benchmark the performance of an atomic quantum simulator against classical simulation to evaluate regimes of quantum advantage \cite{shaw23}.

Extending this approach to 2D enables exploration of the phase-diagram for the interacting Ising system in square, triangular or honeycomb arrays \cite{ebadi21,scholl21} with hundreds of atoms. By varying the final detuning in a square lattice geometry, it is possible to prepare chequerboard (see Fig.~\ref{fig:sim}(c)), striated or star phases distinguished through differences in their correlation functions under a Fourier transform \cite{ebadi21}. For a two-leg ladder it is possible to prepare and characterise floating phases in addition to commensurate $\Z_3 \times \Z_2$ and $\Z_4$ phases \cite{zhang25}. For atoms placed on the links of a kagome lattice as shown in Fig.~\ref{fig:sim}(d), the Rydberg excitations can be mapped to a dimer-model resulting in emergence of a quantum spin liquid at $1/4$ filling, which can be characterised using strong-operators to probe topological properties \cite{semeghini21}. 

Performing quantum simulation using atoms initially prepared in the Rydberg manifold gives access to an $XY$-Hamiltonian \cite{browaeys20} that has enabled observation of symmetry-protected topological edge-states \cite{leseleuc19a} using the Su–Schrieffer–Heeger (SSH) model, a density-dependent Peierls phase \cite{lienhard20} and programmable bosonic $t-J-V$ models \cite{qiao25}, with protocols developed to extend this approach to observe the spin-1 Haldane phase \cite{mogerle25}. These models connect to realistic Hamiltonians of real materials such as frustrated magnets.

The experiments above exploited only the variation in array geometry to provide control over the Hamiltonian. Additional programmability can be achieved by combining this with use of local control fields to apply an independent light shift $\Delta_i$ on each site. This local addressing can be used for controllable state preparation to realise large GHZ states \cite{omran19}, exploration of facilitation dynamics and quantum thermalisation in kinetically constrained models \cite{zhao25}, implementing fast, arbitrary local control of interacting trimers \cite{bornet24}, and measurement of the out-of-time-order correlator (OTOC) to reveal anomalous information scrambling in an atomic chain \cite{xiang24,liang25}. Applying random disorder patterns can further enable investigations of quantum coarsening and collective dynamics in 2D systems \cite{manovitz25}.

In general these quantum simulations consider only the internal degrees of freedom in the atoms. However, the tight confinement of the atomic tweezers combined with techniques for sideband cooling to prepare atoms in the motional ground state \cite{kaufman12} mean it is possible to encode problems in both spin and motional degrees of freedom, using the gradient of the dipole-dipole interaction to create a force which results in a vibronic coupling. This has been proposed as a method to perform quantum simulations of molecular dynamics on neutral atom arrays \cite{euchner25}.

\subsection{Graph Optimisation on Neutral Atom Arrays}\label{sec:graph}
\begin{figure}[t]
\centering
\includegraphics[width=12cm]{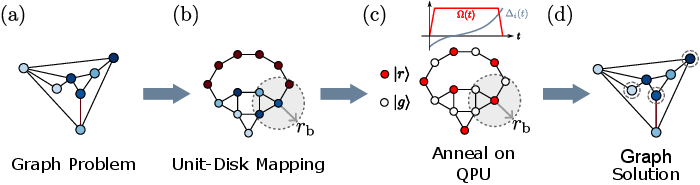}
\caption{Graph optimisation on neutral atom arrays. Starting from an initial graph problem (a), the first step is to embed the problem graph on a unit-disk (UD) graph as shown in (b) such that connected vertices are represented by atoms spaced within a blockade radius. ($r<r_\mathrm{b}$) Additional atoms (shown in dark red) may be required to implement longer-range connections in the original graph. (c) Using the UD embedding, quantum annealing can be performed on a neutral atom quantum processing unit (QPU) to design time-dependent control fields $\Omega(t),\Delta_i(t)$ that adiabatically prepare the atoms in the ground state using closed-loop feedback with the resulting readout measurements mapped to binary variables $\vert g\rangle=0$ and $\vert r \rangle=1$. (d) The ground state solution is then mapped back to give the solution of the target graph problem. \label{fig:mis}}
\end{figure}

The procedures developed for performing quantum simulation on neutral atom arrays can be readily extended to the regime of solving graph optimisation problems \cite{kim23,dalyac24,wurtz24}, whereby instead of exploring the quantum states of the many body system, the annealing process is used to prepare ground states that encode the solution to classical graph optimisation problems as shown in Fig.~\ref{fig:mis}.

The native graph problem implemented on neutral atom arrays is finding the Maximum Independent Set (MIS) on a unit-disk (UD) graph \cite{pichler18,pichler18a}, which is a known NP-hard problem instance \cite{clark90}. A UD graph is one in which only vertices within a given distance from each other are connected by an edge. This can be natively mapped to neutral atom systems using the Rydberg blockade mechanism \cite{jaksch00} which prevents more than a single Rydberg excitation occurring within a radius $r_\mathrm{b}$. For a graph $G(V,E)$ defined as a set of vertices $V$ connected by edges $E$ , the MIS problem is defined as finding the maximum number of vertices that can be selected such that no adjacent (edge-connected) vertices are included. Introducing the binary variables $n_i=\{0,1\}$ to indicate whether vertex $i$ belongs to the UD-MIS solution, the classical cost function that must be minimised is
\begin{equation}
H_\mathrm{MIS}=-\sum_{i\in V} n_j+\sum_{(i,j)\in E} U_{ij}n_in_j,\label{eq:MIS}
\end{equation}
where the first term rewards inclusion of vertices, whilst the second term acts to penalize inclusion of vertices connected by an edge using a large positive term $U_{ij}>1$. Comparison of Eq.~\ref{eq:MIS} to the native Ising Hamiltonian for the Rydberg system in Eq.~\ref{eq:ryd} shows that in the limit $\Omega\rightarrow0$ and $V(a)>\Delta>0$ with edge-connected atoms spaced $a<r_\mathrm{b}$, it is possible to adiabatically prepare the atomic system in the UD-MIS solution using only global control of the Rydberg excitation laser (see Fig.~\ref{fig:mis}(c)).

Experimental demonstrations of solving UD-MIS on 2D graphs with hundreds of qubits validate the feasibility of this approach using both annealing and quantum approximate optimisation algorithm (QAOA) methods \cite{ebadi22,kim24}, with further improvements possible in overcoming limitations of flat energy-landscapes by introducing local control fields during evolution \cite{cain23} or using quench protocols to overcome super-exponentially closing gaps in the energy spectrum of the graphs \cite{schiffer23}. However, recent analysis shows for the native Kings-graph classical solvers are able to find solutions to graphs with thousands vertices in minutes \cite{andrist23}, making UD-MIS a challenging classical problem for which to achieve quantum advantage in the near-term, with previous analysis predicting a cross-over requiring graphs with over 8,000 atoms \cite{serret20}. This target becomes even more challenging when combining the optimisation process with state-of-the-art classical reduction algorithms that seek to take an input graph $G$ and reduce the problem into one or more hard kernels $\mathcal{K}$ which represent a smaller subset of graph nodes. Real graph instances featuring up to 19,000 nodes can be reduced to kernels with only tens to hundreds of nodes, bringing them back in the regime of efficient classical solvers for small kernels \cite{schuetz25a}. This reduction approach has inspired a recent hybrid method combining classical exact solvers with kernelisation methods that leverage samples from a neutral atom device to provide exact MIS solutions \cite{schuetz25}. Including longer-range couplings makes the native Rydberg coupling graph more robust to kernelisation methods \cite{kombe26}, but with work required to develop generic embedding protocols for these regimes.

Given the challenges in achieving quantum advantage in MIS, an alternative approach is to explore Maximum Weighted Independent Set (MWIS) problems, whereby now each vertex gains a weight $w_i$, providing an increased number of parameters in the problem definition. The classical cost function for MWIS becomes
\begin{equation}
H_\mathrm{MWIS}=-\sum_{i\in V} w_in_i+\sum_{(i,j)\in E} U_{ij}n_in_j,\label{eq:MWIS}
\end{equation}
with $U_{ij}>\mathrm{max}_k(w_k)$. Node weights can be implemented experimentally using a spatial light modulator to apply local light-shifts to control the on-site detuning of the Rydberg state $\Delta_i\propto w_i$ by adjusting the relative power of the light-shift on each atom \cite{oliveira25}. In this case the annealing protocol is modified to combine the global detuning of the Rydberg laser with the total power in the light-shift beam that illuminates the spatial light modulator, adding only a single control parameter to control the adiabatic evolution and making it possible to efficiently solve weighted graph instances \cite{oliveira25}.

For many graph problems of interest, it is necessary to map the original problem onto the native UD connectivity offered by the neutral atom platforms. This process, known as graph embedding, enables realisation of non-planar graphs as well as extensions to a larger class of problems including QUBO or integer optimisation problems. A large number of embedding strategies have been proposed and demonstrated, including use of \textit{quantum wires} formed by links of ancilla atoms in 2D \cite{kim22} or routed through neighboring trapping planes in 3D to implement non-local graph connectivity \cite{dalyac23}. This approach has been used to demonstrate embedded integer factorisation \cite{park24} and quadratic unconstrained binary optimisation (QUBO) problems \cite{byun24} onto atom arrays using only global control by mapping to UD-MIS. When mapping to UD-MWIS with local control, weighted quantum wires can be used to connect cliques of 2, 3, or 4 atoms, providing efficient embedding when considering graphs with sparse long-range connectivity as well as enabling native embedding of QUBO problems using wire weights to implement edge couplings \cite{oliveira26}. 

A more generic approach to embedding involves the use of \textit{gadgets} \cite{nguyen23,lanthaler23}, small geometric constructs that can be used as building blocks for constructing larger graphs realising arbitrary non-local connections between logical vertices. The advantage of building gadgets using UD-MWIS is that it enables use of $O(N^2)$ atoms compared to $O(N^6)$ for UD-MIS for a graph problem with $N$ vertices, with the method proposed by Nguyen \textit{et al.} achieving an encoding bounded by $4N^2$ \cite{nguyen23}. These approaches work due to the functional completeness of planar Rydberg blockade structures, enabling construction of arbitrary Hilbert spaces that can be characterised by local constraints in the product basis \cite{stastny23}, however the introduction of additional qubits can lead to challenges in interpreting the final measured output states in the presence of errors \cite{correc25}. More complex gadgets can also be used to implement higher-order constraints \cite{byun24a}.

So far we have focused on problems mapped onto two-level qubits. For integer-optimisation problems, such as minimum vertex graph colouring, the corresponding embedding becomes expensive for $k$-colours on $N$ vertices a minimum of $kN$ qubits is required. Instead, it is possible to exploit the multi-level nature of the Rydberg states and explore qudit based encodings. In this regime, instead we consider the case of using only $N$ atoms with $k$-Rydberg levels, reducing the required atom count by $k$. This provides an efficient embedding for minimum-vertex graph colouring, with theoretical work showing annealing on UD graphs with $k=4$ to validate the approach \cite{angkhanawin25}.

\subsection{Quantum Annealing for Materials Sciences and Quantum Chemistry}
Application of quantum annealing to material science and quantum chemistry problems have been explored in a wide range of contexts, for example crystal structure prediction \cite{gusev23}, chemical reaction networks \cite{mizuno24}, and path-integral molecular dynamics \cite{fiorentino26}. Below, we describe some specific example problems.

Molecular docking, the study of how a given molecule binds with a larger biological target such as protein, has an important application in the context of drug discovery. For biologically relevant molecules, mapping this continuous problem onto a discrete MWIS problem results in extremely large graph sizes, which must be simplified when embedding on real quantum hardware resulting in limitations in accuracy \cite{pandey22,yanagisawa24,triuzzi25}. This has motivated a new workflow developed by Garrigues \etal{} \cite{garrigues25} which allows construction of non-unit disk graphs from realistic molecular problems, which are then broken down into a series of unit-disk MWIS subgraphs compatible with embedding on neutral atom arrays using a quantum heuristic which was benchmarked to outperform greedy algorithms and provide optimal solutions for graph instances with up to 540 nodes. 

A related problem in drug discovery is finding the equilibrium water solvent molecules configurations within proteins. D'Arcangelo \etal{} ~\cite{darcangelo24} map this problem onto an anti-ferromagnetic Ising model compatible with a neutral atom annealer. In this work the continuous 3D Reference Interaction Site Model (3D-RISM) which provides a continuous map of the oxygen atom density inside the protein is mapped to a discrete set of centre-of-mass positions with Gaussian widths. The continuous information can be reconstructed by weighting the sum of Gaussian distributions based on discrete outputs from the quantum annealer. Proof of concept for real proteins are presented using samples of a few tens of atoms, but with a major advantage of the quantum approach compared to current classical methods that it naturally imposes the constraint that two water molecules cannot neighbour eachother, an effect currently added into the classical algorithms.

\jp{For thermodynamic sampling of materials, Camino \etal{} explore quantum annealing using neutral-atom quantum computers for nitrogen-doped graphene \cite{camino26}. In this approach, formation energies calculated using Density Functional Theory (DFT) are mapped onto a Rydberg-atom Hamiltonian, where the on-site energy is controlled through the global laser detuning and pairwise interactions are encoded through the separation between neutral atoms. As the interaction energies obtained from DFT are beyond those directly accessible on current hardware, the authors introduce an energy rescaling that allows the configurations sampled by the annealer to be interpreted in terms of a corresponding effective temperature and chemical potential. The approach is demonstrated on the QuEra Aquila device \cite{aquila23} using a 28-site graphene nanoflake, where the results are benchmarked against exhaustive enumeration, before being extended to a 78-site system with more than \(10^{21}\) relevant configurations. For this larger system, comparison with Monte Carlo sampling shows that the quantum annealer preferentially explores low-energy configurations, whilst varying the neutral-atom separation provides a means of controlling the effective sampling temperature.}


For more generic material science and quantum chemistry problems that can be mapped to intermediate QUBOs, the resulting graphs typically feature many non-local long-range connections, with a worst-case being an all-to-all connected QUBO that requires the full $4N^2$ scaling in atom number. Whilst this presents a significant barrier to near-term usage in a regime competitive with current classical calculation methods, new approaches to graph embedding with longer-range interactions \cite{kombe26} along with increases in atom number may yield routes to future quantum advantage with neutral atom annealers.

\section{Digital Quantum Computing with Neutral Atoms}\label{sec:qec}
\subsection{Fault-tolerant logical operations}

In parallel to the rapid progress in neutral atom technologies for performing analogue quantum simulation and optimisation, there have been major advances in capabilities for performing digital quantum computing. Early demonstrations of two-qubit gates reported limited gate fidelities, due in part to technical noise in the Rydberg lasers used for excitation \cite{isenhower10,leseleuc18}. However, the combination of new techniques for laser stabilisation to suppress phase noise \cite{levine18,levine19} with new time-optimal gate protocols has recently enabled two-qubit fidelities to reach 99.5\% \cite{evered23,ma23}. These fidelities are already competitive with superconducting circuit platforms \cite{jiang25}, with prospects for achieving fidelities exceeding 99.9\% \cite{saffman16} to approach the performance of large-scale trapped ion systems \cite{ransford25}.

\begin{figure}[t]
\centering
\includegraphics[width=10cm]{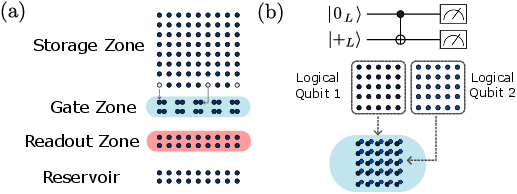}
\caption{(a) Zonal architecture use for scalable digital quantum computation introduced by \cite{bluvstein24,bluvstein25}. The array field of view is divided into four zones - a storage zone for data qubits, a gate zone in which atom pairs placed close together undergo high-fidelity two-qubit gate operations, a readout zone enabling local detection and mid-circuit readout and a reservoir zone for replacing atoms lost during computation. Qubits can be moved in real-time using mobile tweezers providing routes to fully-connected quantum hardware. (b) This platform allows efficient transverse gate operations between logical qubit blocks by using multiple moving tweezers simultaneously to combine $5\times5$ blocks of atoms that each encode a single logical qubit in the gate-zone and applying parallel two-qubit gate operations between the atom pairs \cite{bluvstein24}.\label{fig:zone}}
\end{figure}

Alongside improved fidelities, the ability to dynamically reconfigure atomic qubits using a steerable tweezer whilst maintaining coherence has overcome limitations of nearest-neighbour connectivity experienced by many other platforms \cite{bluvstein22}, enabling development of zonal architectures formed with spatially separated regions in the array for qubit storage, parallel gate operations and qubit readout \cite{bluvstein22,bluvstein24,bluvstein25}, as shown schematically in Fig.~\ref{fig:zone}(a). For scaling up operations based on movement, new efficient hardware-aware compilers have been developed to optimise routing and scheduling of moves during computation \cite{tan24,tan25}.

For large scale quantum computing it is essential to be able to apply quantum error correction \cite{roffe19}, in which logical qubits are encoded across a number of physical qubits, with additional ancilla qubits used to perform measurements that can detect (and correct) errors as the computation progresses. For neutral atom arrays, the development of high-quality mid-circuit readout \cite{ma23,graham23,bluvstein24,bluvstein25} combined with high-quality gates has enabled demonstrations of logical qubit operations \cite{bluvstein24}, magic state distillation \cite{rodriguez25} and initial quantum error correction codes on these platforms \cite{bluvstein25, zhang25, reichardt25}, where the ability to spatially reconfigure and readout many atoms in parallel enables efficient transversal gate operations, as illustrated in Fig.~\ref{fig:zone}(b). Similar performance has been obtained using a static configuration where, instead of moving atoms, local addressing beams are used to sequentially apply two-qubit gates between neighbouring qubits \cite{graham22}. The native movement along row and columns of the array is also well adapted for realising quantum low-density parity check (qLDPC) codes \cite{xu24} which can provide much higher efficiency encodings of logical qubits on a given number of physical qubits than the equivalent surface code, at the cost of increased complexity for logical operations \cite{xu25}.

Whilst these advances have shown that neutral atom platforms are highly scalable and offer many near-term advantages for fault-tolerant computing, there remain a number of challenges unique to this approach \cite{saffman16,bluvstein25}. The first relates to relative timescales, where whilst gates occur on microsecond timescales, measurement typically takes 0.5-10~ms \cite{bluvstein25, zhang25, reichardt25}. This comparatively slow readout rate limits the clock cycle, requiring development of new techniques for fast readout such as the use of optical cavities \cite{deist22}, protocols for using additional ancilla qubits for fast parallel readout \cite{petrosyan24,corlett25}, or integration with low-latency single-photon avalanche detector (SPAD) arrays \cite{zeng26}. The second is dealing with atom loss during computation, which occurs due to spontaneous decay from the Rydberg state or finite vacuum lifetimes \cite{saffman16}. This has inspired new loss-aware protocols for error correction which seek to reduce the number of cycles over which a given atom is used before checking for loss, either by using swap-based operations to periodically exchange the role of data and ancilla qubits \cite{baranes25,yu25a} or measurement-based quantum computing protocols \cite{yu25}. Loss and heating during atom motion is predominantly limited by intermodulation effects in the RF based acousto-optical deflectors \cite{bluvstein25}. Whilst this can be suppressed through careful choice of atom motion, in future this could be eliminated by using novel optical devices that enable fast mapping from frequency to position in two-dimensions \cite{wei26,deters26}. Finally, due to qubit loss the ability to perform real-time atom replacement to extend the duration of the logical computation is also essential, which has motivated development of new experimental hardware able to implement continuous reloading to maintain arrays with over 1000 atoms over many hours \cite{gyger24,norcia24,muniz25,chiu25}.

Current experiments have predominantly focused on single-species architectures due to simplification of laser requirements and overall experimental complexity. However, in these single-species platforms a challenge arises from light scattering during mid-circuit measurement being resonant with data qubits, combined with limited performance when scaling from two- to multi-qubit gate operations \cite{levine19}. Attempts to perform in-situ mid-circuit measurement using microwave shelving shows this is possible, albeit with relatively slow readout times \cite{graham23}, whilst the zonal approach requires movement of ancilla qubits to a readout zone that is spatially separated from the data qubits, and applying a shielding light-shift beam to hide data qubits during measurement to shift the excited state resonances \cite{hu25,bluvstein25}. Whilst this works well, it imposes a challenge in requiring large, high power lasers for the shielding beams and there is a finite off-resonant scattering that can enhance idle errors. 

An alternative solution is to use dual-species arrays with a large wavelength separation between atomic species, which have been demonstrated using Rb and Cs at room temperature to allow in-situ non-destructive imaging \cite{singh22}. Use of optimised interspecies Rydberg interactions  enables high fidelity two- and multi-qubit gates with reduced cross-talk \cite{beterov15,ireland24}, as well as potential to perform accelerated readout by performing controlled operations with ancilla qubits to reduce the total imaging time below 1~ms \cite{corlett25}. This also permits use of multi-qubit gates for performing efficient syndrome extraction, for example in the surface code enabling four $CZ$ operations to be replaced with a pair of sequential three-qubit $CZ^2$ gates \cite{pecorari25}. Complementary approaches have focused on use of alkaline-earth qubits (Yb,Sr) which feature long-lived metastable states. By encoding qubits in the metastable states, most errors can be mapped to back to the low-lying ground states enabling erasure conversion using fast imaging pulses which can detect an error when imaging atoms in the ground state and replace them with qubits initialised into the metastable qubit basis \cite{wu22}, which is also compatible with mid-circuit measurements \cite{norcia23,lis23}.

\subsection{Prospects for Near-term Applications}
Advances in neutral atom computing platforms reveal a promising route towards scalable and fault-tolerant operation, with conservative resource analysis estimating that using the transversal logical encodings and parallel operations can provide significant improvements in space-time efficiency, reducing the time-to-solution to the order of 6 days for solving 2048-bit RSA factorisation using hundreds of thousands of atoms \cite{zhou25,cain26}. These resource estimates assume 1~ms QEC cycles, faster than current platforms but  with recent demonstrations of 15~$\mu$s readout and low-latency feedforward using SPADs \cite{zeng26} remains a realistic timescale.

In future, these large-scale platforms offering all-to-all connectivity between logical qubits will be ideally suited for tackling large-scale digital algorithms targeting demanding material science and quantum chemistry problems introduced above, where even simple material models require thousands of gates on hundreds to thousands of qubits \cite{clinton24}.  However, near-term systems currently with few logical qubits can provide useful testbeds for exploring algorithm performance, such as recent implementation of an Anderson Impurity Model ground state solver on a pair of logical qubits \cite{chung25}. Improvements in gate performance has also already opened new pathways for performing direct digital quantum simulation of programmable Hamiltonians, with demonstrations of probing topological features of the Kitaev model using local fermionic encoding by applying Floquet engineering \cite{evered25}. This approach can also be extended to perform Trotterised evolution of a sparsely-coupled $XY$ model for creating metrologically useful entangled states \cite{kuriyattil25}.

The pathway towards quantum advantage in materials science and quantum chemistry applications therefore remains challenging, but has motivated development of new algorithms targeting reduced gate and qubit overheads. Babbush \etal{} for example introduce a new low-depth algorithm for quantum simulation of the electronic structure problem that reduces the total overhead from simulating an Hamiltonian with $\mathcal{O}(N^4)$ terms to a lower bound depth of $\mathcal{O}(N)$ \cite{babbush18}. This team subsequently developed fault-tolerant protocols for simulating the condensed-phase correlated electrons via Trotterization on a surface code \cite{kivlichan20}. Kan \etal{} consider resource-optimized fault-tolerant simulation of the Fermi-Hubbard model as an early application, and demonstrate only an order of magnitude increase in Toffoli gate count for extending this to more realistic models of high-temperature superconductors including cuprates and iron-pnictide \cite{kan25}. In the context of catalysis, new algorithms offering an order of magnitude improvement have been developed to more efficiently calculate the electronic energy of key intermediates and transition states of its catalytic cycle \cite{vonburg21}. In these applications, even with improved algorithms these have significant resource requirements of $10^5$ physical qubits and $>10^6$ T-gates putting them far beyond the performance of current devices.

One alternative route is to consider applications involving modular quantum processors, where a number of smaller devices can be networked together. This has motivated a new distributed unitary selective coupled cluster algorithm for quantum chemistry applications \cite{xue26} which requires only modest connectivity, with examples presented for performing Trotterized dynamics of a (H$_4$)$_3$ chain created from clusters of 24 qubits.

\section{Fermionic Quantum Computing}\label{sec:fermion}
For many material science and quantum chemistry applications, the underlying physical system is comprised of fermions, which possess the property of being anti-symmetric under exchange. This has motivated development of programmable quantum simulators for fermions such as quantum gas microscopes \cite{haller15,cheuk15} that can now reach ultracold temperatures around 0.05 of the Fermi temperature \cite{xu25fermion} for studies of the Hubbard model. However, when using digital quantum computers for simulating fermionic systems, a key limitation arises in that whilst standard qubit systems are comprised of effective spin-$1/2$ particles, these do not replicate the intrinsic fermionic symmetries. Current approaches exploit encodings such as Jordan-Wigner to map the spin exchanges onto the parity of qubit strings, increasing the overhead and making long-range or many-body couplings expensive to implement.

An alternative paradigm is that of fermionic quantum computing, whereby the qubits are encoded on fermionic particles that natively respect the spin-exchange statistics. This approach has motivated development of topologically protected Majorana-based fermionic quantum computing \cite{obrien18}, but can be readily adapted to the neutral atom platform by use of fermionic atomic species, and the combination of long-range Rydberg interactions with tunnelling gates between atoms in neighbouring optical tweezers \cite{gonzalez-cuadra22}. Theoretical work has further developed protocols for fermionic error correction in tweezer arrrays \cite{ott25}. This fermionic approach provides a significant reduction in qubit overhead when compared to alternative methods based on Jordan-Wigner encodings, with recent analysis finding an exponential reduction in circuit depth for Trotterized time evolution under the Hamiltonian of crystalline materials \cite{schuckert25}. In this work authors reduce the gate depth for a single time-step from $\mathcal{O}(N)$ on a qubit-based algorithm to $\mathcal{O}(\log(N))$ on a fault-tolerant fermionic processor when simulating dynamics on $N$-lattice sites, whilst also using fewer ancilla atoms.

Whilst fermionic tweezer and spin-coupling introduces new experimental challenges, recent demonstrations of fermions in optical lattices illustrate the viability of applying controlled tunnel-coupled gate operations \cite{bojovic25} and have motivated development of new programmable hybrid systems that combine tweezers and lattices as a credible route to fault-tolerant fermionic quantum processors \cite{jain26}.

\section{Conclusion}
Neutral atom quantum computing provide highly scalable platforms that are well adapted for performing programmable quantum simulation and analogue quantum optimisation, as well as implementing large-scale quantum error correction. Recent advances targeting increased array sizes, improved readout times and increased gate fidelities have permitted rapid progress towards fault-tolerant operation of logical qubits, as well as novel protocols for embedding graph optimisation problems relevant across a wide range of applications \cite{wurtz24}.

For the demanding requirements of quantum chemistry and material science problems, there remains significant work to reach the regime of thousands of logical qubits required for utility scale hardware \cite{menssen26}. However, the rapid rate of progress both on hardware and algorithms, combined with exploration of new neutral atom architectures and demonstrations of continuous reloading means fault-tolerant platforms offering tens to hundreds of logical qubits will be available within the timescale of a few years, opening new avenues for simulating programmable Hamiltonians of complex materials, and testing new algorithms targeting applications in quantum chemistry.

%

\end{document}